\documentclass[pra,showpacs,preprintnumbers,amsmath,amssymb]{revtex4}
\usepackage{bm}
\usepackage{amsmath}
\usepackage[dvips]{graphicx}
\begin{document}
\bibliographystyle{apsrev}

\title{Vacuum polarization radiative correction to the parity violating
electron scattering on heavy nuclei}

\author{A. I. Milstein}
\email[Email:]{A.I.Milstein@inp.nsk.su} \affiliation{Budker
Institute of Nuclear Physics, 630090 Novosibirsk, Russia}
\author{O. P. Sushkov}
\email[Email:]{sushkov@phys.unsw.edu.au} \affiliation{School of
Physics, University of New South Wales, Sydney 2052, Australia}

\date{\today}
\begin{abstract}
The effect of vacuum polarization on the parity violating
asymmetry in the elastic electron-nucleus scattering is
considered. Calculations are performed in the high-energy
approximation with an exact account for the electric field of the
nucleus. It is shown that the radiative correction to the parity
violating asymmetry is  logarithmically enhanced and the value of the
correction is about -1\%.
\end{abstract}
  \maketitle

Experimental  investigations of the proton $\rho_p(r)$ and neutron
$\rho_n(r)$ densities in heavy nuclei are very important for
testing of various nuclear models. The densities are also important
for many applications like, for example, precise calculations of
atomic parity nonconservation. While the charge distribution (proton
density) is known pretty well mainly due to data on the elastic electron
scattering, the determination of the neutron density with a good accuracy
is a very complicated problem \cite{Batty1989}.
It was suggested by Donnelly {\it et al} \cite{Donnelly1989}  to
use parity violation (PV) in the electron-nucleus scattering to determine
the neutron distribution. The idea behind this suggestion is that the
mediator of the weak interaction, Z-boson, interacts mainly with neutrons.
The parity violating asymmetry is defined as
\begin{equation}\label{A}
{\cal A}_{PV}=
\frac{d\sigma_+/d\Omega-d\sigma_-/d\Omega}{d\sigma_+/d\Omega+d\sigma_-/d\Omega}\,
,
\end{equation}
where $d\sigma_+/d\Omega$ and $d\sigma_-/d\Omega$ are cross
sections for scattering of right-handed and left-handed electrons,
respectively. The asymmetry can be measured experimentally and in
order to determine the neutron distribution one needs to calculate
the asymmetry at a given distribution. The scattering potential is
of the form
\begin{equation}\label{V}
V(r)=V_C(r)+\gamma_5\,A(r)\, ,\quad
A(r)=\frac{G_F}{2\sqrt{2}}\,\varrho_W(r)\, .
\end{equation}
Here $V_C(r)$ is the Coulomb potential of the nucleus,
$G_F=1.16639\times 10^{-5}GeV^{-2}$ is the Fermi constant,
$\gamma_5$ is the  Dirac matrix, and $\varrho_W(r)$ is the density
of the weak charge
\begin{equation}\label{rhow}
\varrho_W(r)=-N\varrho_n(r)+(1-4\sin^2\theta_W) Z\varrho_p(r)\, ,
\end{equation}
where  $\theta_W$ is the Weinberg angle ($\sin^2\theta_W\approx
0.23$),  $N$ and $Z$ are numbers of protons and neutrons,
respectively. The densities are normalized as $\int d\bm r\varrho_{p,\,n}(r)=1$.
In this work we consider a high energy scattering, so  the electron mass can be
neglected compared to energy. In this limit the cross sections $d\sigma_+/d\Omega$ and
$d\sigma_-/d\Omega$ correspond to scattering in potentials
 $V_C(r)+A(r)$ and $V_C(r)-A(r)$,
respectively. The PV asymmetry is due to interference between the
 Coulomb amplitude and the
amplitude caused by the axial potential $A(r)$. Since neutrons dominate
in the weak-charge density $\varrho_W(r)$  (\ref{rhow}), the
asymmetry ${\cal A}_{PV}$ is very sensitive to the shape of
$\varrho_n(r)$. Though the value of ${\cal A}_{PV}$ is very small
and its measurement is a very complex problem, there is a proposal \cite{prop} for
the corresponding experiment.

Detailed theoretical investigation of the asymmetry
 ${\cal A}_{PV}$ has been performed by Horowits \cite{Hor98},
Vretenar {\it et al} \cite{Vret00}, and by  Horowits {\it et al} \cite{Hor01}.
It has been shown that ${\cal A}_{PV}$ is very
sensitive to the small difference between $\varrho_p(r)$ and
$\varrho_n(r)$. It has been also shown that the Coulomb distortions of the electron
wave functions significantly modify the asymmetry as compared to
that obtained in the plane-wave approximation. In this case radiative corrections can
be enhanced. The corrections have not been considered before.
In the present work we calculate the vacuum polarization radiative
 correction to ${\cal A}_{PV}$.
 We found that similar to radiative corrections to atomic parity
violation \cite{Mil0,Milw} the correction we consider here is logarithmically
enhanced. The logarithm is $\ln(\lambda_C/r_0)$ , where  $r_0 \approx 1.5 \ Z^{1/3}
\ fm $ is the nuclear radius and $\lambda_C$ is the electron Compton wavelength.

\begin{figure}[ht]
\vspace{40pt}
\centering
\includegraphics[height=200pt,keepaspectratio=true]{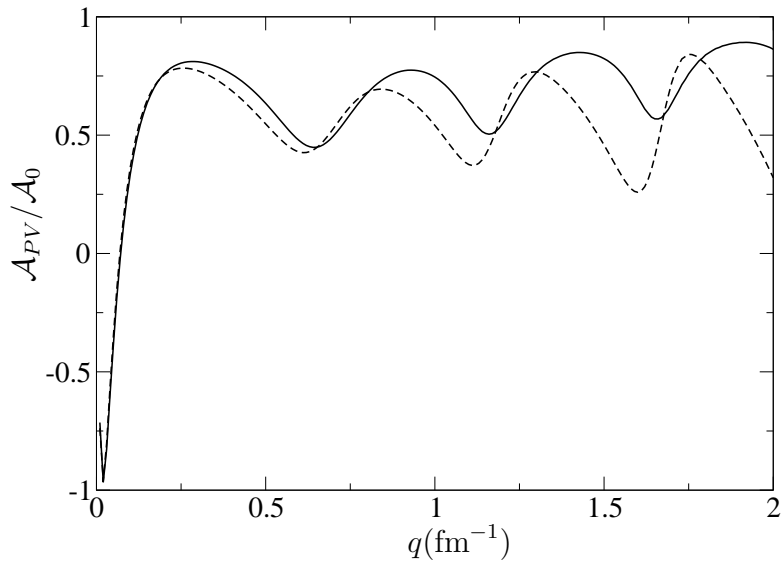}
\begin{picture}(0,0)(0,0)
 \put(-145,-7){\large $q(\mbox{fm}^{-1})$}
 \put(-295,95){\rotatebox{90}{\large${\cal A}_{PV}/{\cal A}_0$}}
 \end{picture}
\caption{\it Dependence of the ratio ${\cal A}_{PV}/{\cal A}_0$
for Pb (Z=82) on the momentum transfer \ $q$.  The solid line
corresponds to $\varrho_n(r)=\varrho_p(r)$ and the dashed line
corresponds to $\varrho_n(r)=\lambda^3\varrho_p(\lambda r)$ with
$\lambda=0.95$. The reference Born approximation asymmetry ${\cal
A}_0$ is given by Eq. (\ref{A0})} \label{Fig1}
\end{figure}

It is well known that in the leading $Z\alpha$  approximation
($\alpha=1/137$ is the fine-structure constant), the vacuum
polarization results in the Uehling potential \cite{Ueh}. At
$r_0\ll r \ll \lambda_C$, this potential is of the form
$V(r)\approx 2Z\alpha^2[\ln(r/\lambda_C)+C+5/6]/(3\pi r)$, where
$C\approx 0.577$ is the Euler constant. Account of corrections
higher in $Z\alpha$ leads to a modification of the constant:
 $C \to C + 0.092Z^2\alpha^2+...$,
see Ref. \cite{Mil}. This correction is small and can be
neglected even for $Z\alpha \sim 1$. For our purpose it is
necessary to calculate the vacuum polarization potential at $r\sim
r_0$ where the presented above simple logarithmic approximation is not
sufficient. Nevertheless the conclusion about smallness of the
$Z\alpha$-corrections to the Uehling potential remains valid at $r\sim r_0$
and hence we will use the Uehling potential in our calculations.
In Refs. \cite{Hor98,Vret00,Hor01}, the scattering amplitude has
been obtained as a result of summation over partial waves.
We do not use a partial wave analysis, instead we use the high-energy
small-angle approximation. The scattering amplitude in this approximation reads
\begin{eqnarray}
\label{f}
 &&f=-\frac{ip}{2\pi}\int d\bm \rho
 \exp[-i\bm q\bm\rho]
 \left\{\exp[-i\chi(\rho)]-1\right\}\ ,\nonumber\\
&&\chi(\rho)=\int_{-\infty}^{\infty}dz\, V(r)\ , \
r=\sqrt{z^2+\rho^2}\, ,
\end{eqnarray}
where $\bm q=\bm p_2-\bm p_1$ is the momentum transfer, z-axis is directed
along the vector $\bm p=(\bm p_2+\bm p_1)/2$ , and $\bm\rho$ is a
two-dimensional vector orthogonal to $\bm p$. Here $\bm p_1$ and
$\bm p_2$ is the initial and the final electron momentum, respectively.
Validity of Eq. (\ref{f}) has been discussed in numerous works, see book \cite{ube}
and references therein.
At first, this formula has been derived using wave functions in the eikonal
approximation. Then, Olsen {\it et al} \cite{OMW57} have demonstrated that the range of validity
is wider: the semiclassical approximation is sufficient to justify Eq. (\ref{f})
even if the eikonal approximation is not valid.
Corrections to Eq. (\ref{f}) are considered in
 Ref. \cite{LMS00}, see also Ref. \cite{ABS}.
For high-energy small-angle scattering the semiclassical approximation
and hence formula (\ref{f}) is well justified and we will use it.
From Eqs. (\ref{A}) and (\ref{f}), we obtain
\begin{eqnarray}
\label{Aeik}
 &&{\cal A}_{PV}=2\frac{\mbox{Re}\left[f_C^*f_W\right]}{\left
 |f_C\right|^2}\, \nonumber\\
&& f_C=-\frac{ip}{2\pi}\int d\bm \rho\,\exp[-i\bm q\bm\rho]
 \left\{\exp[-i\chi_C(\rho)]-1\right\}\,\nonumber\\
&& f_W=-\frac{p\,G_F}{4\sqrt{2}\pi}\int d\bm \rho\,\exp[-i\bm
q\bm\rho
 -i\chi_C(\rho)]\int_{-\infty}^{\infty}dz\,\varrho_W(r)\,\nonumber\\
&&\chi_C(\rho)=\int_{-\infty}^{\infty}dz\, V_C(r)\, ,
\end{eqnarray}
To simplify further numerical integrations, it is convenient to express
$\chi_C(\rho)$ in terms of  $\varrho_p$,
\begin{eqnarray}
\label{chi}
 &&\chi(\rho)=2Z\alpha\,\Phi(\rho)\, ,\nonumber\\
&&\Phi(\rho)=\ln(\rho/L)+4\pi \int_\rho^\infty dr\, r
\varrho_p(r)\left[r\,\ln\left(\frac{r+\sqrt{r^2-\rho^2}}{\rho}\right)-
\sqrt{r^2-\rho^2}\right]\, .
\end{eqnarray}
Here $L\gg r_0$ is an arbitrary constant, the cross sections and the
asymmetry are independent of the constant. The function $\Phi(\rho)$ has the
following properties:
\begin{equation}
\label{Phi}
 \Phi(\rho)= \left\{
\begin{array}{c}
4\pi \int_0^\infty dr\, r^2 \varrho_p(r)\ln(2r/L)-1\, , \quad
\mbox{at}\, \rho\to 0\, ,\\
\ln(\rho/L)\, , \quad
\mbox{at}\, \rho\gg r_0\,. \\
\end{array}
\right.
\end{equation}
Taking in Eq. (\ref{Aeik}) the integral over the angle of the
vector $\bm \rho$ and using Eq. (\ref{chi}), we obtain
\begin{eqnarray}
\label{fCW}
 && f_C=\frac{2Z\alpha p}{q^2}\left\{4\pi\int_0^\infty
 d\rho\,\rho\,
 J_0(q\rho)\exp[-2iZ\alpha\Phi(\rho)]\int_{\rho}^{\infty}\frac{dr\,r\,\varrho_p(r)}
 {\sqrt{r^2-\rho^2}}\right.\,\nonumber\\
&&\left.-2iZ\alpha\int_0^\infty
 \frac{d\rho}{\rho}\,J_0(q\rho)
 \exp[-2iZ\alpha\Phi(\rho)]\,\tilde\Phi^2(\rho)\right\}\, ,\nonumber\\
&& f_W=-\frac{p\,G_F}{\sqrt{2}}\int_0^\infty
d\rho\,\rho\,J_0(q\rho)\exp[-2iZ\alpha\Phi(\rho)]
\int_{\rho}^{\infty}\frac{dr\,r\,\varrho_W(r)}
{\sqrt{r^2-\rho^2}}\, ,
\end{eqnarray}
where
 $$\tilde\Phi(\rho)=\rho\frac{\partial\Phi(\rho)}{\partial\rho}
 =1-4\pi\int_{\rho}^{\infty}dr\,r\varrho_p(r)
 \sqrt{r^2-\rho^2}\, ,$$
 and $J_0(x)$ is the Bessel function.
Amplitudes in form (\ref{fCW}) are very convenient for numerical integration
due to fast convergence of the integrals.
In the Born approximation (plane wave approximation)
the amplitudes and the asymmetry are
\begin{eqnarray}\label{Born}
 && f_C^B=\frac{2Z\alpha p}{q^2}F_p(q)\quad ,\quad
 f_W^B=-\frac{p\,G_FF_W(q)}{4\pi\sqrt{2}}\,,\nonumber\\
&&{\cal A}_{PV}^B= -\frac{q^2G_FF_W(q)}{4\pi\sqrt{2} Z\alpha
F_p(q)}=\frac{q^2G_F
}{4\pi\sqrt{2}\alpha}\left[\frac{NF_n(q)}{ZF_p(q)}+4\sin^2\theta_W-1\right]\,
.
\end{eqnarray}
Here $F_p(q)$ , $F_n(q)$ , and $F_W(q)$ are Fourier transforms
(form factors) of corresponding densities.
If neutron and proton distributions coincide, $\varrho_n(r)=\varrho_p(r)$, then
the Born approximation asymmetry (\ref{Born})  is given by
\begin{equation}\label{A0}
{\cal A}_0=\frac{G_F
q^2}{4\pi\sqrt{2}\alpha}\left[\frac{N}{Z}+4\sin^2\theta_W-1\right]\,
.\end{equation}
We will use this asymmetry as a reference point.  Figure \ref{Fig1} shows
dependence of the ratio ${\cal A}_{PV}/{\cal A}_0$ on the momentum transfer $q$ for
Pb (Z=82)  at $\varrho_n(r)=\varrho_p(r)$ (solid curve) and at
$\varrho_n(r)=0.95^3\varrho_p(0.95 r)$  (dashed curve).
For $\varrho_p(r)$ we use the three parameter Fermi charge density \cite{Hor98}:
\begin{eqnarray}
\label{rhop}
 &&\varrho_p(r)=\varrho_0\frac{1+w(r/r_0)^2}
 {1+\exp[(r-r_0)/a]}\quad
 ,\nonumber\\
&& r_0=6.4\, \mbox{fm}\, ,\quad a=0.54\, \mbox{fm}\, ,\quad
w=0.32\, ,\quad \varrho_0=6.95\times10^{-4}\, \mbox{fm}^{-3}
\end{eqnarray}
Plots in Fig. \ref{Fig1} are in excellent agreement with results of the
partial wave analysis \cite{Hor98}.
The figure clearly demonstrates high  sensitivity to the
 difference between $\varrho_n(r)$
and $\varrho_p(r)$ as well as importance of the Coulomb distortion.
The distortion is especially important at small $q$. At $q\ll 1/r_0$ the
Coulomb and the weak amplitudes are of the form
\begin{eqnarray}
\label{as}
 && f_C=\frac{2Z\alpha p}
 {q^2}\left(\frac{qL}{2}\right)^{2iZ\alpha}
 \frac{\Gamma(1-iZ\alpha)}{\Gamma(1+iZ\alpha)}\quad ,\nonumber\\
&& f_W=-\frac{p\,G_F}{\sqrt{2}}\int_0^\infty
d\rho\,\rho\,\exp[-2iZ\alpha\Phi(\rho)]\int_{\rho}^{\infty}\frac{dr\,r\,\varrho_W(r)}
{\sqrt{r^2-\rho^2}}\, .
\end{eqnarray}
So at $q\to 0$ the weak amplitude $f_W$ is independent
 of $q$ while phase of the Coulomb amplitude
$f_C$ strongly depends on $q$. This explains oscillations
 of ${\cal A}_{PV}$ at $q \to 0$.

 Let us consider now influence of vacuum polarization on the parity violating
asymmetry. The  charge density $\varrho_{vp}(r)$ induced by
polarization of the electron-positron vacuum and expressed in
units of the elementary charge $|e|$ is of the form, see e.g. the
review paper \cite{BR82}
\begin{eqnarray}
\label{ICD}
 &&  \varrho_{vp}(r)= \frac{Z\alpha}{3\pi}\int_1^\infty
d\tau\,\sqrt{\frac{\tau-1}{\tau}}\,\frac{(\tau+1/2)}{\tau^2}\nonumber\\
&&\times\left\{\varrho_p(r)-\frac{m^2\tau}{\pi}\int d\bm R\,
\frac{\exp(-2m\sqrt{\tau}\,|\bm r-\bm R|)}{|\bm r-\bm
R|}\,\varrho_p(R)\right\} \, .
\end{eqnarray}
This charge density corresponds to the Uehling potential \cite{Ueh}.
For small distances, $r\ll \lambda_C$,  expression (\ref{ICD}) can be
transformed to
\begin{eqnarray}
\label{ICD1}
 &&  \varrho_{vp}(r)\to \frac{2Z\alpha}{3\pi}\left\{
\left[\ln(\lambda_C/r)-C-5/6)\right]\,\varrho_p(r) \right.\nonumber\\
&&+\frac{1}{r} \int_0^r dR\,\varrho_p(R)-\frac{1}{2}
\int_0^\infty dR\, \frac{\varrho_p(R)}{r+R}\nonumber\\
&&\left.-\frac{1}{2}
\int_0^{2r}dR\,\frac{\varrho_p(R)-\varrho_p(r)}{|R-r|}
-\frac{1}{2} \int_{2r}^\infty dR\,\frac{\varrho_p(R)}{R-r}\right\}
 \, .
\end{eqnarray}
Note that Eq. (\ref{ICD1}) is not singular at $r=0$, all the divergent terms
are canceled out. Outside of nucleus, at $r\gg r_0$,
Eq.({\ref{ICD}) is equivalent to the well known formula for the induced charge
density of a point-like nucleus
\begin{eqnarray}
\label{ICD2}
 &&  \varrho_{vp}(r)= -\frac{2Z\alpha \,m^2}{3\pi^2 r}\int_1^\infty
dx\,\sqrt{x^2-1}\,\left(1+\frac{1}{2x^2}\right)\exp(-2mrx) \, .
\end{eqnarray}
At $\lambda_C\gg r\gg r_0$ it gives
\begin{equation}
\label{ICD3} \varrho_{vp}(r)= -\frac{Z\alpha }{6\pi^2 r^3} \, .
\end{equation}
A direct numerical integration of (\ref{ICD}) is straightforward.
In  Fig. \ref{Fig2} we show by solid line the vacuum polarization
charge density $\varrho_{vp}(r)$ for Pb. For comparison the dashed
line presents the proton density (\ref{rhop}). We remind that
$\varrho_{p}(r)$ is normalized on unity, $\int d\bm r
\varrho_{p}(r)=1$.
\begin{figure}[ht]
\vspace{10pt}
\centering
\includegraphics[height=170pt,keepaspectratio=true]{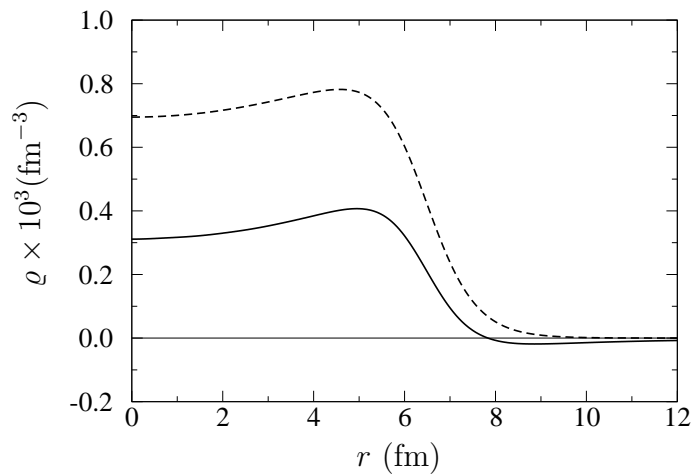}
\begin{picture}(0,0)(0,0)
 \put(-130,-3){\large $r$ (fm)}
 \put(-260,65){\rotatebox{90}{\large$\varrho\times 10^3(\mbox{fm}^{-3})$}}
 \end{picture}
\caption{\it Solid line: the vacuum polarization charge density
$\varrho_{vp}(r)$ for Pb in units  $10^{-3}|e|/fm^3$. Dashed line: the
proton density $\varrho_{p}(r)$ in units $10^{-3}\mbox{fm}^{-3}$.
The proton density is normalized to unity. }
 \label{Fig2}
\vspace{10pt}
\end{figure}
We see that in the vicinity of the nucleus the vacuum
polarization charge density is practically proportional to the
proton density. With logarithmic accuracy this fact immediately
follows from Eq. (\ref{ICD1})
\begin{eqnarray}
\label{ICD4} \varrho_{vp}(r)\sim \frac{2Z\alpha}{3\pi}
\ln\left(\frac{\lambda_C}{r_0}\right)\varrho_p(r)  \, .
\end{eqnarray}
It is interesting that inside a sphere of radius $8\,  fm$
the vacuum polarization charge
is rather big and equals to $0.47|e|$. So one can say that the electric charge
of the Pb nucleus is 82.47. Certainly the excess
 charge 0.47 is compensated
by the negative charge -0.47 distributed between $r_0$ and
$\lambda_C$, see Eq. (\ref{ICD3}). Using Eqs. (\ref{ICD1}),
(\ref{ICD2}) and (\ref{fCW}), it is easy to calculate the
asymmetry ${\cal A}_{tot}$ that account for the effect of vacuum
polarization. One should replace $\varrho_{p}(r)$ in (\ref{fCW})
to $\varrho_{p}(r)+\frac{1}{Z}\varrho_{vp}(r)$. Note that it is not necessary
to account for the effect of vacuum polarization on $\varrho_W(r)$
because the corresponding correction contains the additional
suppression factor $1-4\sin^2\theta_W$ in the electron-$Z^0$
vector vertex in the electron loop. With account of the radiative
correction the parity violating asymmetry is reduced,
 ${\cal A}_{PV}\to{\cal A}_{PV}(1+\Delta)$.
 Value of the vacuum polarization radiative correction $\Delta$ in per cent
is plotted in Fig. ~\ref{Fig3} versus the momentum transfer $q$.
\begin{figure}[ht]
\vspace{50pt} \centering
\includegraphics[height=180pt,keepaspectratio=true]{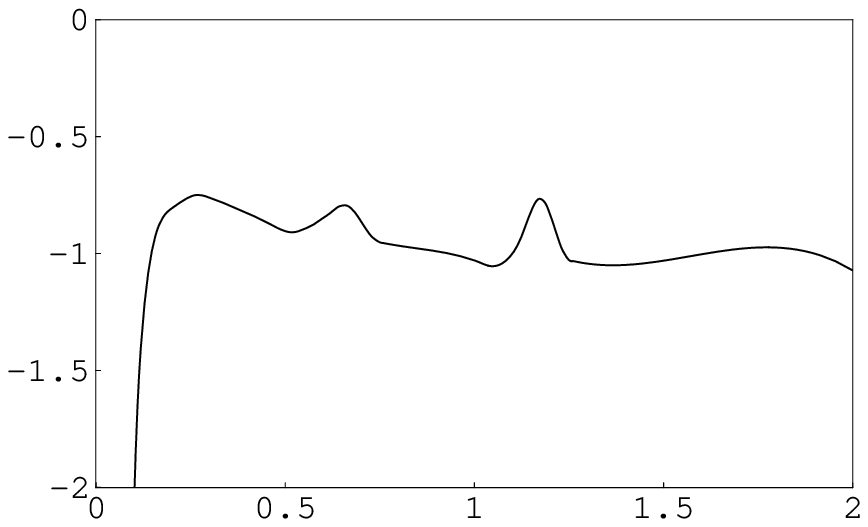}
\begin{picture}(0,0)(0,0)
 \put(-145,-7){\large $q(\mbox{fm}^{-1})$}
 \put(-310,95){\rotatebox{90}{\large$\Delta(\%)$}}
 \end{picture}
\vspace{50pt} \caption{\it The vacuum polarization radiative
correction to the parity violation asymmetry (\%) versus the
momentum transfer. } \label{Fig3} \vspace{10pt}
\end{figure}
In the interesting range of momenta, $q=0.2\div 2fm^{-1}$, the
radiative correction is about -1\%. A similar value one can expect
in the Born approximation. In this approximation only the
denominator $F_p(q)$ in Eq.(\ref{Born}) is influenced by the
correction. Therefore, due to (\ref{ICD4}) the radiative
correction is
\begin{equation}
\label{dB}
\Delta^{(Born)} \sim -\frac{2\alpha}{3\pi}
\ln\left(\frac{\lambda_C}{r_0}\right) \approx -0.7\%
\end{equation}

We have considered the vacuum polarization correction to the parity
 violating asymmetry. There are also electron self-energy
and vertex corrections corresponding to radiation and absorption
of virtual photons, and to radiation of real photons.  It is well
known that these corrections to the scattering amplitude are very
big due to double logarithms. For example the vertex correction to
the Born scattering cross section is \cite{BLP}
\begin{equation}
\label{1}
\Delta_{\sigma}=\exp\left(-\frac{2\alpha}{\pi}\ln\frac{-q^2}{m^2}
\ln\frac{E}{\Omega}\right)-1 \ .
\end{equation}
Here $q$ is the momentum transfer, $E$ and $m$ are the electron
energy and mass, respectively, and $\Omega$ is resolution with
respect to bremsstrahlung. For  conditions of proposal \cite{prop}
($E=850MeV$, $q=90MeV$, $\Omega=4MeV$) the correction is -23\%
(!). However, in the Born approximation the vertex correction is
canceled out in the parity violating asymmetry ${\cal A}_{PV}$.
Certainly the Born
 approximation is not sufficient to
describe electron scattering on Pb, and  a calculation of  the vertex correction to
${\cal A}_{PV}$ in this case is an open and a very hard problem.

Concluding we have calculated the effect of vacuum polarization on
the parity violating asymmetry in the elastic electron-nucleus
scattering. The vacuum polarization radiative correction  is
logarithmically enhanced and the value of the correction is about
-1\%.

A.I.M. gratefully acknowledges the School of Physics at the
University of New South Wales for warm hospitality and financial
support during a visit. The work  was also supported by RFBR Grant
No. 03-02-16510.

\end{document}